\documentclass[twocolumn]{aastex7}

\usepackage{enumitem}
\usepackage{graphicx}
\usepackage{subfig}

\newcommand{\fudesig}{FUOr-Sgr 1816-1951}
\newcommand{\objir}{WTP 19aabuix}
\newcommand{\objopt}{ZTF 24abgjhjq}

\received{17 June 2026}

\submitjournal{ApJL}

\begin{document}

\title{V7995 Sgr: A New FU Orionis Accretion Outburst Near NGC 6589/6590}

\author{Lynne A. Hillenbrand}
\affiliation{Department of Astronomy, MC 249-17, California Institute of Technology, Pasadena, CA 91125; USA}
\email{lah@astro.caltech.edu}

\author[orcid=0000-0002-8989-0542]{Kishalay De} 
\affiliation{Department of Astronomy and Columbia Astrophysics Laboratory, Columbia University, New York, NY, USA}
\affiliation{Center for Computational Astrophysics, Flatiron Institute, New York, NY, USA}
\email{kd3038@columbia.edu}

\author[0000-0002-9540-853X]{Adolfo S. Carvalho} 
\affiliation{Center for Astrophysics, Harvard University, 60 Garden St., Cambridge, MA, 02138, USA}
\email{adolfo.carvalho@cfa.harvard.edu}

\author[0000-0003-2686-9241]{Dan Stern}  
\affiliation{Jet Propulsion Laboratory, California Institute of Technology, 4800 Oak Grove Drive, Pasadena, CA 91109; USA}
\email{Daniel.K.Stern@jpl.nasa.gov}

\author{Evan Portnoi}  
\affiliation{Department of Astronomy, MC 249-17, California Institute of Technology, Pasadena, CA 91125; USA}
\email{evanrossportnoi@gmail.com}

\author[0000-0001-6627-9903]{Nicholas Earley} 
\affiliation{Cahill Center for Astrophysics, California Institute of Technology, Pasadena, CA 91125, USA}
\email{earley@caltech.edu}

\begin{abstract}
We announce a new FU Orionis type outburst that reached peak brightness in late 2024,
following a steep 4.6 month photometric rise of -2.85 mag in the $r$ band. 
This rapid brightening at all wavelengths was preceeded in the infrared
by a much shallower rise over 4 years. 
The progenitor object was an unstudied young stellar object having a flat-spectrum type 
spectral energy distribution, and extended nebulosity.  
We present multi-wavelength lightcurves covering the photometric low-state, the outburst,
and early post-outburst epochs.  
Optical imaging shows a concurrent brightening of the extended nebular environment.
We also present follow-up optical/near-infrared spectroscopy
taken 1.5 years after the inferred photometric peak.  
The spectra confirm an FU Ori type outburst. The outburst source exhibits
a mixed-temperature absorption spectrum, formed in an accretion disk,
and it shows several line species with blueshifted absorption profiles that are formed in a strong wind.  
\end{abstract}


\section{Introduction } \label{sec:intro}

Episodic accretion in young stellar objects (YSOs) occurs over a 
wide range of amplitudes and across a wide range of timescales 
\citep{fischer2023}.  The most extreme events are dubbed FU Orionis-type outbursts,
named after the prototype FU Ori \citep{herbig1966}. 
FU Ori eruptions are marked by dramatic brightening, typically reaching $\sim$4-6 mag in the optical 
and $\sim$1-3 mag in the infrared, with rise times spanning months to a few years. 
They are widely interpreted as the consequence of disk instabilities that trigger 
substantial increases in the accretion rate, 
boosting the accretion luminosity by factors of $\sim10^{2-4}$ \citep{Hartmann1996}. 

In an outburst state, optical and near-infrared emission is dominated by the hot inner
accretion disk atmosphere, rather than the stellar photosphere  \citep{kenyon1988,zhu2009}.
FU Ori accretion outbursts are sustained on long timescales of decades to potentially centuries.
In addition to the implication of such episodic accretion for stellar mass assembly, 
FU Ori events also induce heating of the inner protoplanetary disk, which can affect
the thermal structure and chemistry over a wide range of disk radii during the epoch of planet formation
\citep{laznevoi2025, houge2023}.


\section{Background on V7995 Sgr}

Our source of interest is located at R.A. = 18:16:40.03, Decl. = -19:51:56.4. 
It is about 1\arcdeg.6 south of the Galactic Plane and about 5 degrees in projection 
from famous star forming regions like M8 (the Lagoon Nebula) and M20 (the Trifid Nebula).
This area in Sagittarius is not well-studied, but the source is located 
in a dark region not far in projection from several small reflection nebulae: 
NGC 6590 illuminated by the cluster NGC 6595,
and NGC 6589 illuminated by the star HD 167638 \citep{hubble1922,magakian2003}. 
The object is likely proximate in all three dimensions to these small reflection nebulae. 

Gaia astrometry to the source is unreliable. \cite{bailer-jones2021} derive an uncertain geometric distance 
of 3.2 kpc (2.3-4.3 kpc for the 16th and 84th percentile posteriors).
However, we prefer to adopt the inverse parallax distance of 1310 pc to HD 167638, 
which is known to a few percent \citep{GDR3}.
Consistent with this distance, Bayestar dust extinction maps \citep{green2019} show jumps 
in color along the line of sight corresponding to $A_V \approx 5.0$ mag  
at 1.25 kpc distance, and then around 3 kpc to $A_V \approx 10.3$ mag.  

\begin{figure}[ht]
    \centering
\includegraphics[width=\linewidth]{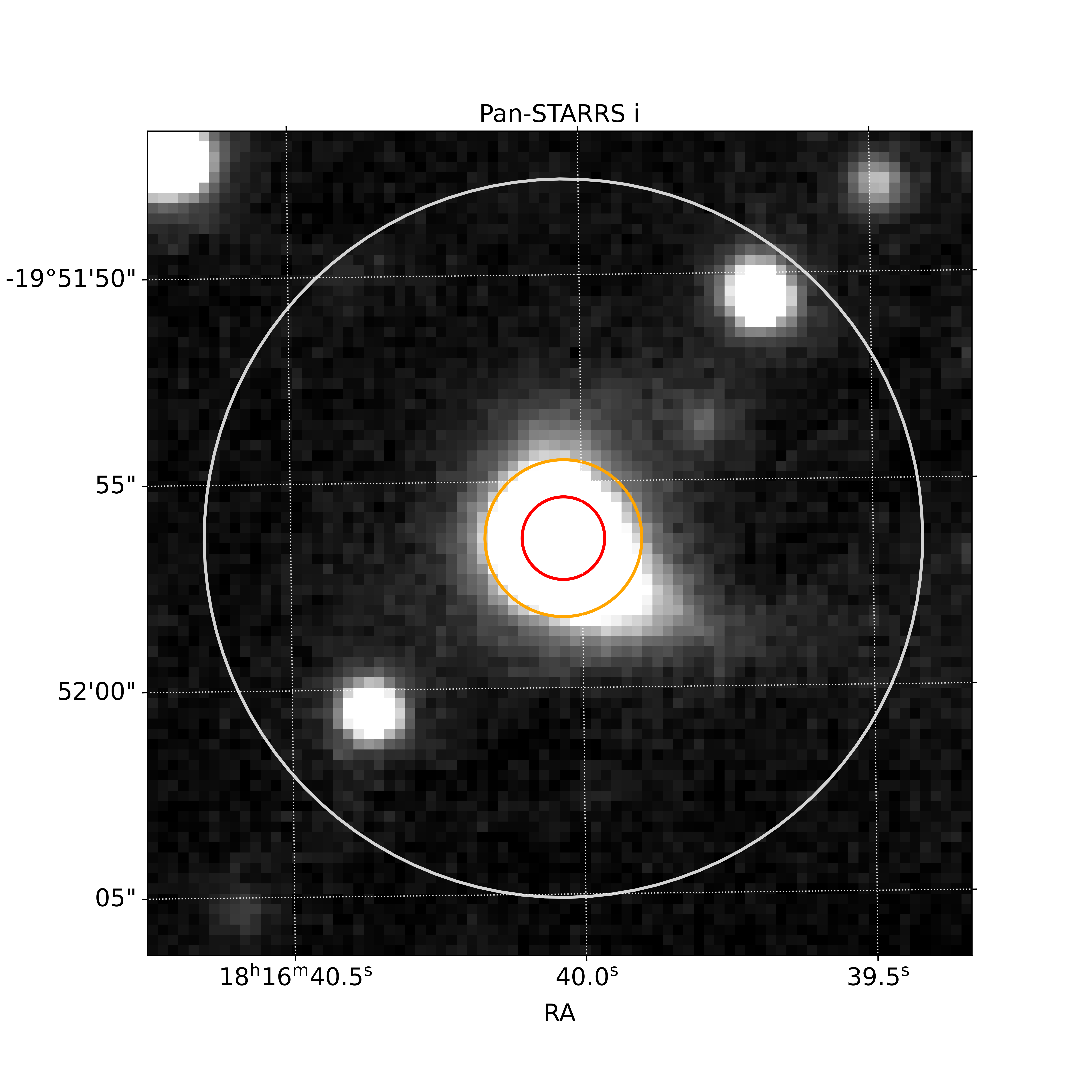}
    \caption{
    PanSTARRS i-band image of the pre-outburst source and environment.
    The object is clearly extended compared to near-by point sources.
    Red, orange, and gray circles show the inscribed radius of a single pixel, 
    or about the 1/2-radius aperture, for the respective ZTF, ATLAS, and Gattini photometry.
    }
    \label{image}
\end{figure}

Information on the pre-outburst source can be found only in large photometric catalogs, 
and in several photometric variability catalogs. 
The source is also present in the large YSO catalog of \cite{marton2016}.
Available pre-outburst optical and infrared photometry indicate the source SED as a flat spectrum type YSO
having spectral index $\alpha_{[2-22~\mu{\rm m}]}=-0.27$. 
It is nebulous in PanSTARRS imaging (Figure~\ref{image}). 

Based on the results presented below, the object has 
received the GCVS\footnote{http://www.sai.msu.su/gcvs/gcvs/}
\citep[General Catalog of Variable Stars;][]{samus2017} designation V7995 Sgr.

\section{Observations and Results} \label{sec:obs}
\subsection{Photometric Outburst Detection} \label{sec:phot}

\begin{figure}[!ht]
    \centering
\includegraphics[width=1.1\linewidth]{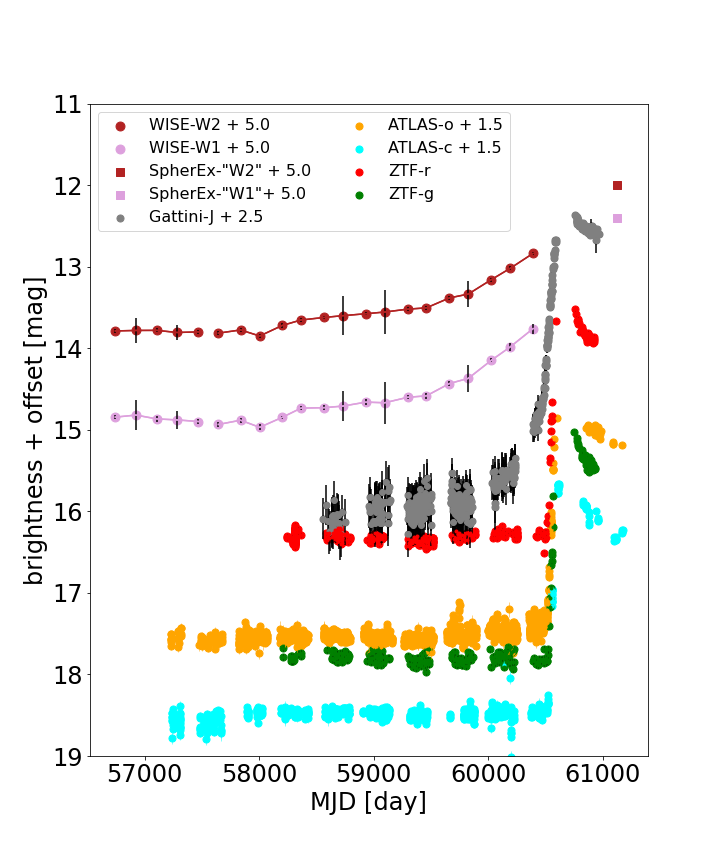}
    \caption{
    Lightcurves in optical ZTF g-band (0.48 $\mu$m), ZTF r-band (0.64 $\mu$m),
    ATLAS c-band (0.53 $\mu$m), and ATLAS o-band (0.69 $\mu$m); 
    near-infrared Gattini $J$-band (1.6 $\mu$m); 
    and mid-infrared NEOWISE W1 (3.6 $\mu$m), W2 (4.5 $\mu$m).
    Shown are observed magnitudes for ZTF, with other bands offset for 
    compactness and clarity.
    }
    \label{lcs}
\end{figure}

The object was initially identified as a potentially interesting lightcurve
in a search of the unWISE \citep{lang2014,meisner2018} data products that result from coaddition of 
WISE (2010; \citealt{wise_overview}) and NEOWISE (2014-2024; \citealt{mainzer2014}) 
data \citep{neowise_data} taken near in time.  For the search, we employed an image-subtraction pipeline \citep{zackay2016,de2020} 
to identify sources with sustained brightness increases.
The identified mid-infrared brightening event was designated as \objir.

Separately, and much later, the source alerted in the ZTF \citep{bellm2019,graham2019} data stream, 
in which it was designated as \objopt\ \citep{patterson2019}. 
We pulled ZTF $g,r$ photometry \citep{masci2019} from IPAC\footnote{https://irsa.ipac.caltech.edu/applications/ztf/}.
We also pulled\footnote{https://fallingstar-data.com} ATLAS \citep[Asteroid Terrestial-impact Last Alert System;][]{atlas}
$c,o$ photometry.
And we recovered the source in the Palomar Gattini InfraRed \cite[PGIR][]{de2020} time series database \citep{gattini_data},  
providing $J$-band photometry.
We corrected the Gattini photometry for two $15^m$ sources within the aperture (Figure~\ref{image}).

\begin{figure}[!ht]
    \centering
\includegraphics[width=1.1\linewidth]{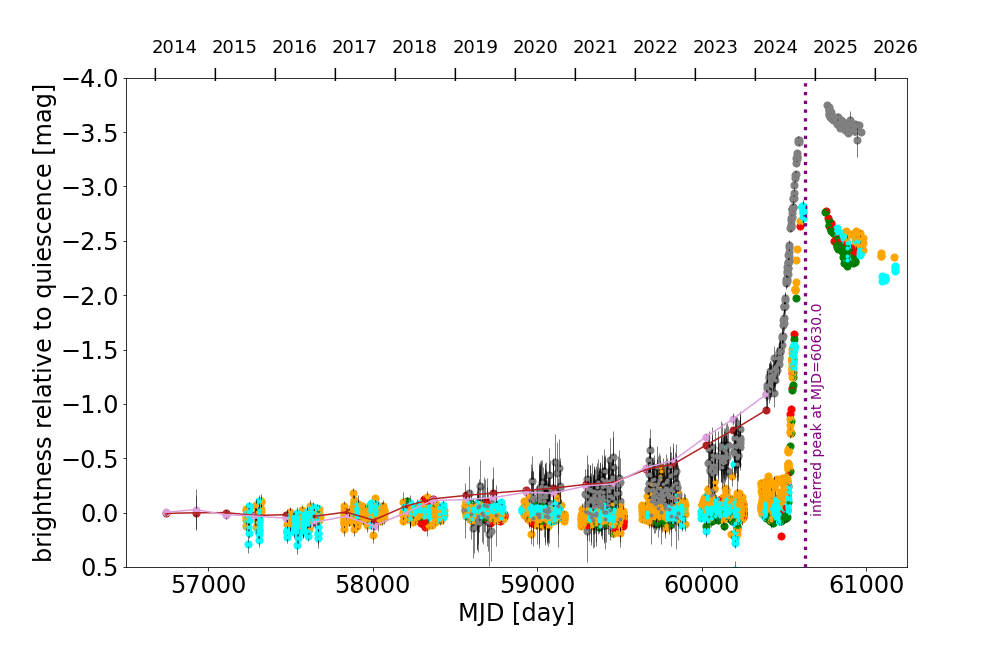}
\includegraphics[width=1.1\linewidth]{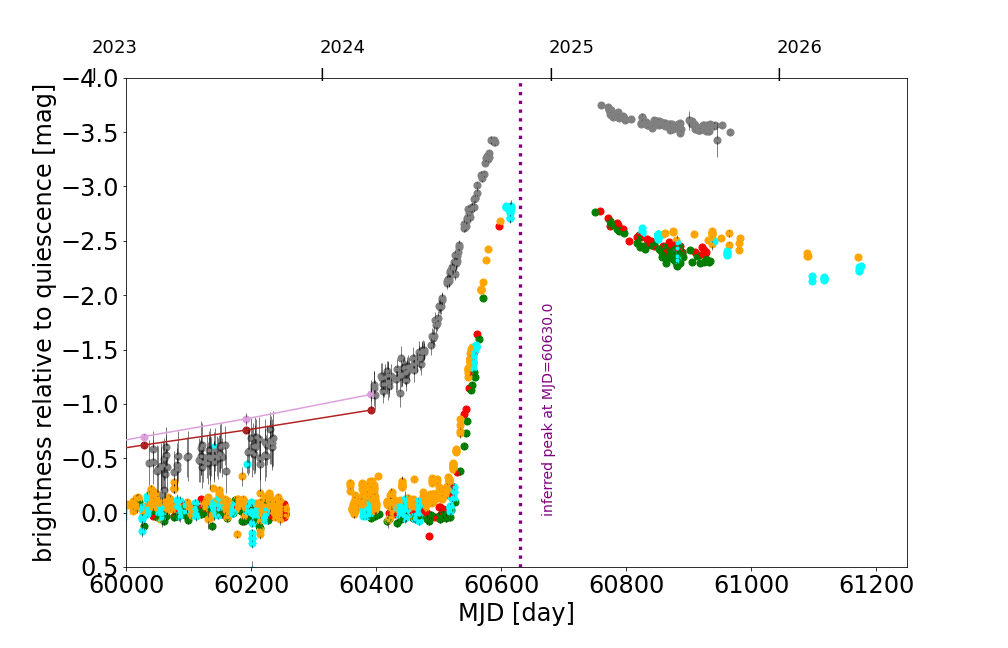}
    \caption{
    The same multi-wavelength lightcurves as in Figure~\ref{lcs} but with the
    magnitudes shifted in each band by their median value at MJD$<$59000, offset to zero.
    }
    \label{lcs_shift}
\end{figure}

The lightcurves are shown in Figures~\ref{lcs} and ~\ref{lcs_shift}. 
We note that there appears to be differences in the post-peak evolution 
between the ATLAS and the ZTF photometry, with the ATLAS points systematically brighter.
We suspect that this is due to a photometric aperture size difference, given the ATLAS 1\arcsec.9 pixels
and the ZTF 1\arcsec.0 pixels which therefore sample different areas of the extended source.

The optical photometric brightening also resulted in changes to the surrounding nebulosity. 
See Appendix for a demonstration of this in the ZTF images.

In the low-state, the source was variable at optical and
near-infrared wavelengths at the 0.2 mag level. 
Figure~\ref{lcs_shift} illustrates the lightcurves offset in each band 
by their median magnitudes before MJD=59000.  These median values are:
$g=17.79$, $c=16.99$, $r=16.30$, $o=16.04$, $J=13.61$, $W1=9.85$, and $W2=8.78$ mag, 
with resulting colors $g-r=1.49$ and $W1-W2$=1.07.

The outburst appears to have brightened first in the mid-infrared,
then in the near-infrared, and finally and also more sharply at optical wavelengths. 

Around MJD=58000 the source began a shallow brightening in $W1$ and $W2$. 
It started a blueing trend in $W1-W2$ by MJD=59000, becoming at least -0.2 mag
bluer. By MJD=60393, the end of the NEOWISE data stream, the source had  
brightened by at least -1.1 mag in $W1$ and -0.9 mag in $W2$. 
There are post-peak measurements from SpherEx \citep{spherex}
for which synthetic photometry can be performed to approximate the $W1$ and $W2$ bands.
These show continued blueing, by another 0.4 mag. 
The bluer-when-brighter color-magnitude trend has fitted slope $W1-W2$ vs $W1$ of $0.14\pm0.02$,
and is distinct from reddening variations. 

In the near-infrared, the $J$-band brightening began around MJD=59000, 
near or slightly after the mid-infrared brightening began,
but years before the optical brightening.

In the optical, the source remained relatively constant until MJD=60500 
when a distinctive outburst began.
The lightcurve shows the sharp rise and then an immediate decay from peak brightness along an exponential.
The impulsive burst itself is red, with color increasing from $g-r=1.5$ in quiescence to 1.8 near the peak.
However, the optical color quickly dropped in the post-peak era back to the pre-outburst value.

All observations in hand miss the photometric peak, 
which is inferred through function fitting. 
After normalizing a composite of the four bands of ZTF and ATLAS photometry 
as described above (Figure~\ref{lcs_shift}),
we estimate that a peak brightness of $r=13.47$ mag occurred on MJD=60630 (2024-11-16).
From the Gattini photometry, we estimate a peak brightness of $J=9.86$ mag occurred on MJD=60677 (2025-01-02).
We also find that the rise time from the quiescent state to lightcurve peak is 4.5 ($J$-band) to 4.6 ($r$-band) months. 
Despite the differences in the derived numbers, there is no evidence that either the peaks or the steep rise times 
are actually different between the optical and near-infrared.
However, the total infrared rise times are indeed much longer. There was an initial shallow rise 
in both the near-infrared and mid-infrared that began about 4 years earlier than 
the later and steeper rise coincident with the optical brightening.

\subsection{Spectroscopic Follow-up} \label{sec:spec}

The outburst source was observed with the Palomar 200" Hale telescope and its
Next Generation Palomar Spectrograph \citep[NGPS;][]{jiang2018},
and with the Keck II telescope and its Near InfraRed Echellette Spectrograph \citep[NIRES;][]{tspecs2004}. 
Both spectra were taken on 2026-05-08 UT.

\begin{figure}[!ht]
    \centering
\includegraphics[width=\linewidth]{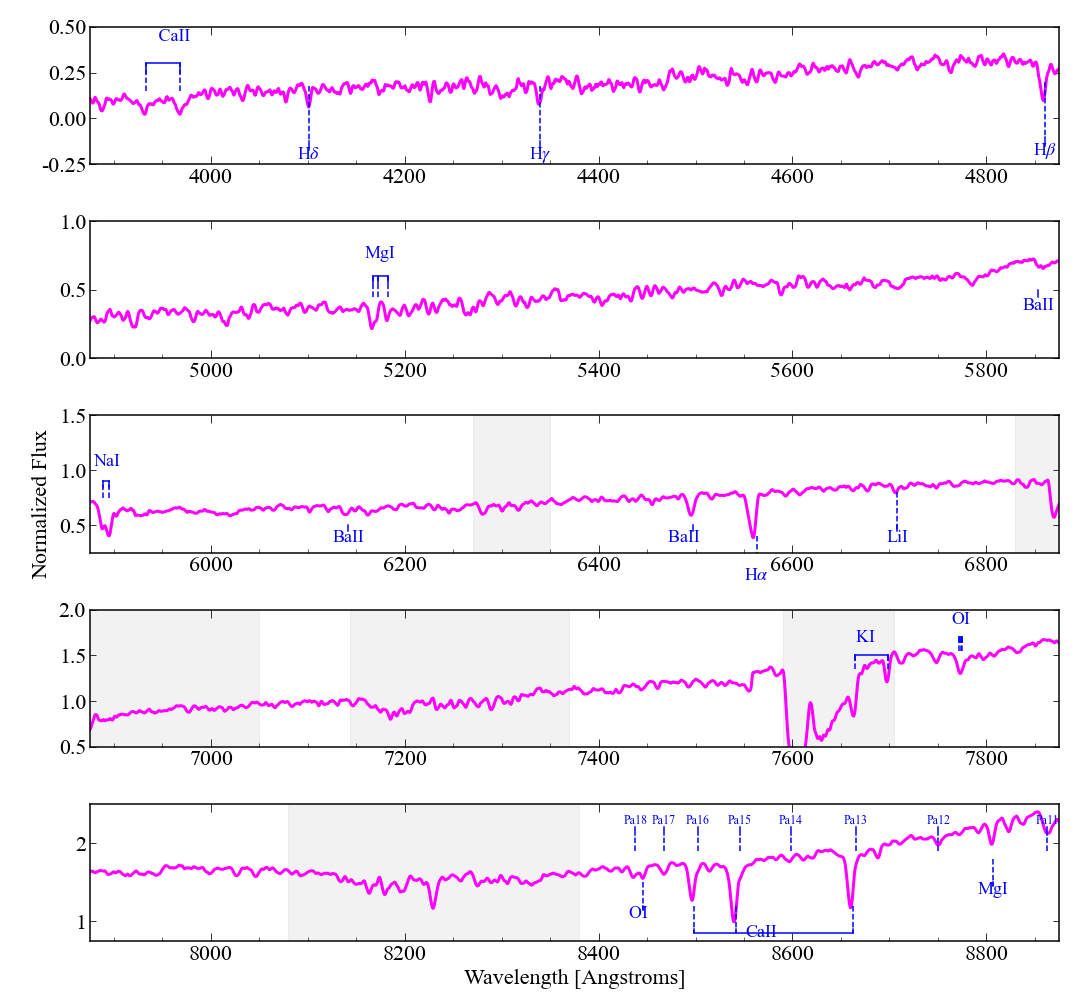}
    \caption{
    Optical spectrum of the outbursting source (magenta).
    Features typically seen in absorption in FU Ori type spectra are marked (blue),
    with moderate blueshifts exhibited in lines of H$\alpha$, H$\beta$, \ion{Ca}{2}.
    Shading (gray) indicates regions of substantial atmospheric contamination.
    }
    \label{optspec}
\end{figure}

In the optical we obtained a single 300 s observation using a 1.5\arcsec\ wide slit oriented at the parallactic angle. The night was clear and the seeing at the time of the observation was about 1.5\arcsec. We processed the data using standard procedures, and flux-calibrated using the sd0C standard star PG0823+546 \citep{massey1990;} obtained the same night.
Figure~\ref{optspec} shows the optical spectrum, normalized by its median flux. 

\begin{figure}[!ht]
    \centering
\includegraphics[width=\linewidth]{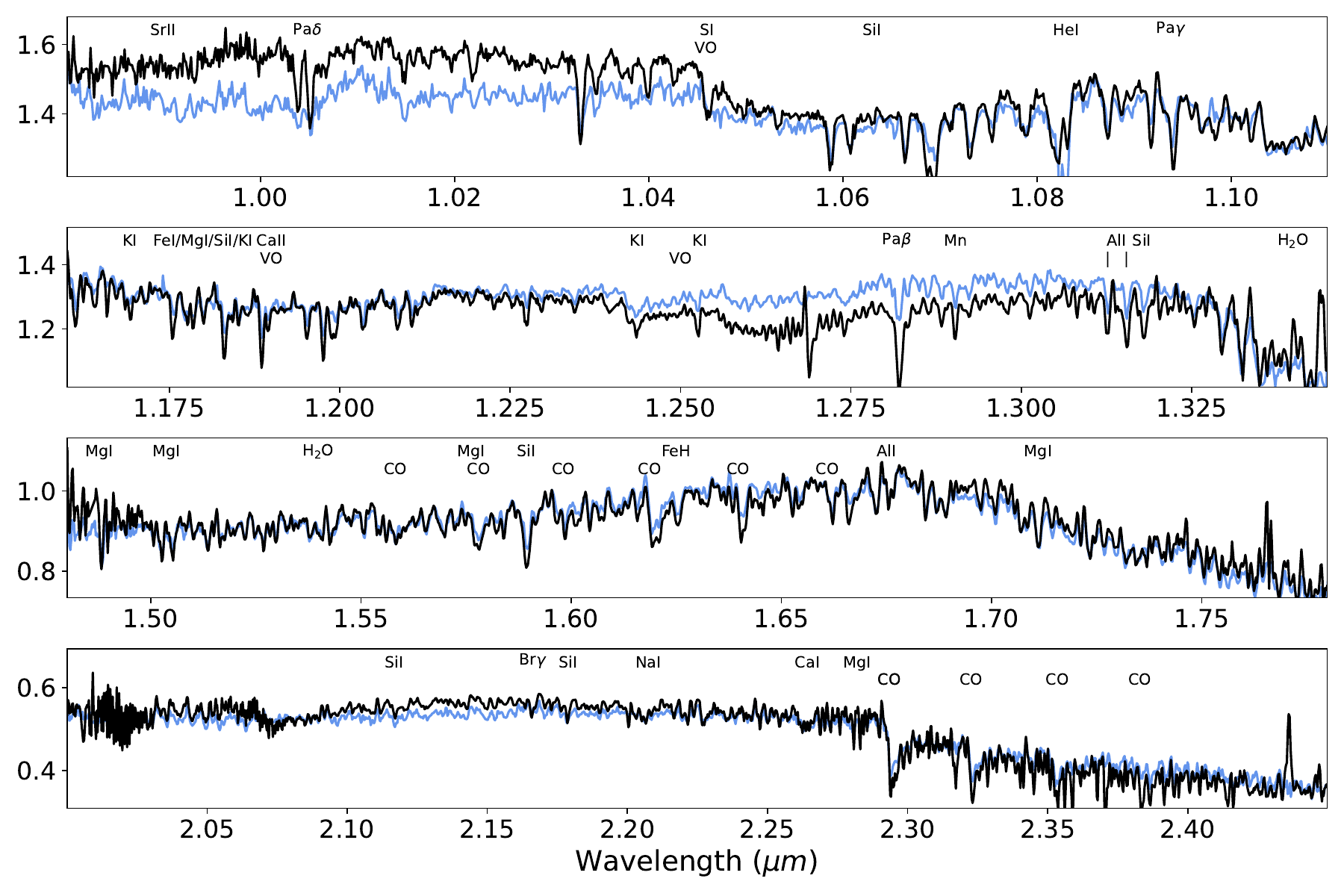}
    \caption{
    Infrared spectrum of the outbursting source (black) 
    compared to the median of 28 spectra of bona fide FU Ori sources (blue) from \cite{portnoi2026}.
    }
    \label{irspec}
\end{figure}

In the infrared, we observed at the parallactic angle using a standard ABBA dithering pattern with 120 s total exposure time. 
The images were processed and the spectra extracted and calibrated using the PypeIt pipeline
\citep{prochaska2020}. The resulting SNR is $>100$ everywhere, and reaches $>500$ in the K-band.
Figure~\ref{irspec} shows the near-infrared spectrum, normalized by its value around $1.78 \mu$m, and
dereddened using the method of \cite{portnoi2026} resulting  in $A_V=4.8\pm0.5$ 

The spectra exhibit the key features of an FU Ori type object, which consist of 
a multi-temperature atomic and molecular spectrum, as expected from an accretion disk,
and blueshifted absorption in certain wind-diagnostic lines.
Our spectra show absorption in the optical from atomic lines of e.g. \ion{Ba}{2}, \ion{Fe}{2}, and \ion{Mg}{1},
and in the infrared from atomic \ion{Na}{1}, \ion{Ca}{1}, \ion{Al}{1}, \ion{Mg}{1}, \ion{Mn}{1}, \ion{Si}{1}, and \ion{Sr}{2}, 
as well as from molecular bands of TiO, H$_2$O, CO, and FeH. 
There is blueshifted absorption in many key wind features, specifically 
H$\alpha$, H$\beta$, \ion{Na}{1}D, the \ion{K}{1} doublet, \ion{O}{1} triplet, 
and \ion{Ca}{2} doublet and triplet in the optical,
and Pa$\beta$, Pa$\delta$, Pa$\gamma$, Br$\gamma$, and \ion{He}{1} 10830 \AA\ in the infrared.

\section{Further Analysis and Discussion}

The steep rise timescale was about 4.6 months, as documented in 
the optical $\Delta r = -2.85$ mag brightness increase and the final parts
of the near-infrared $J$ band brightness increase.  However, this steep rise
to photometric maximum was preceeded in the near- and mid-infrared 
by a much shallow brightness increase over about 4 years,
amounting to $\Delta J = -3.75$ mag and $\Delta W1$ exceeding -2.45 mag.

Infrared brightening that occurs prior to optical brightening has been
interpreted as outside-in propogation of a disk instability
\citep{bae2014,cleaver2023}.  However, the results of \citet{das2026} suggest
that such behavior may be a consequence of temperature evolution 
in the inner accretion disk, caused by the outward expansion of the 
dust sublimation front. 

The near-infrared and mid-infrared amplitudes are larger than typical
for FU Ori outbursts.  However, the rise amplitude in the optical 
is relatively small for an FU Ori. 
The optical outburst lightcurve bears some similarity 
to that of V960 Mon, an FU Ori object that outburst in 2014. 

The -2.85 optical amplitude is comparable to V960 Mon's -2.8 mag \citep{carvalho2023}.
Also similar to V960 Mon is the immediate evolution from peak brightness
into an exponential decline (Figures~\ref{lcs}, and ~\ref{lcs_shift}).
We note that V960 Mon exhibited significant evolution in its lightcurve and spectrum over
the first several years post-peak \citep{carvalho2023,carvalho2024}.

For both sources, the models of \cite{das2026}
with this particular lightcurve morphology indicate that the inner disk reaches the star
for only a brief period around lightcurve peak.  Such impulsive lightcurves that 
rise quickly and transition immediately to exponential decay may be consistent
with fly-by like triggering mechanisms for FU Ori outbursts \citep{vorobyov2021}.

Multi-wavelength follow-up is encouraged, as the spectrum and lightcurves are likely to evolve
over the next few months to year. 
Should the current spectroscopic absorption features and photometric bright-state persist,
following the convention introduced in \cite{hillenbrand2025} the source would be named \fudesig.

\begin{acknowledgements}
This work was supported, in part, by NASA under awards \#80NSSC23K0655 and \#80NSSC24K0663. 
We are grateful to the entire NEOWISE, Gattini, ZTF, and ATLAS teams for making this kind of time domain science possible.
We are also grateful to the observatory staffs at the W.M. Keck and Palomar Observatories for facilitating our spectroscopic observations.
We thank Javier Acevedo Barroso and Anuraag Upadhyayula for participating in the Palomar observations.
Palomar Gattini-IR (PGIR) is generously funded by Caltech, Australian National University, the Mt Cuba Foundation, the Heising Simons Foundation, the Binational Science Foundation. PGIR is a collaborative project among Caltech, Australian National University, University of New South Wales, Columbia University and the Weizmann Institute of Science. 
\end{acknowledgements}

\begin{figure}[ht]
    \centering
\includegraphics[width=\linewidth]{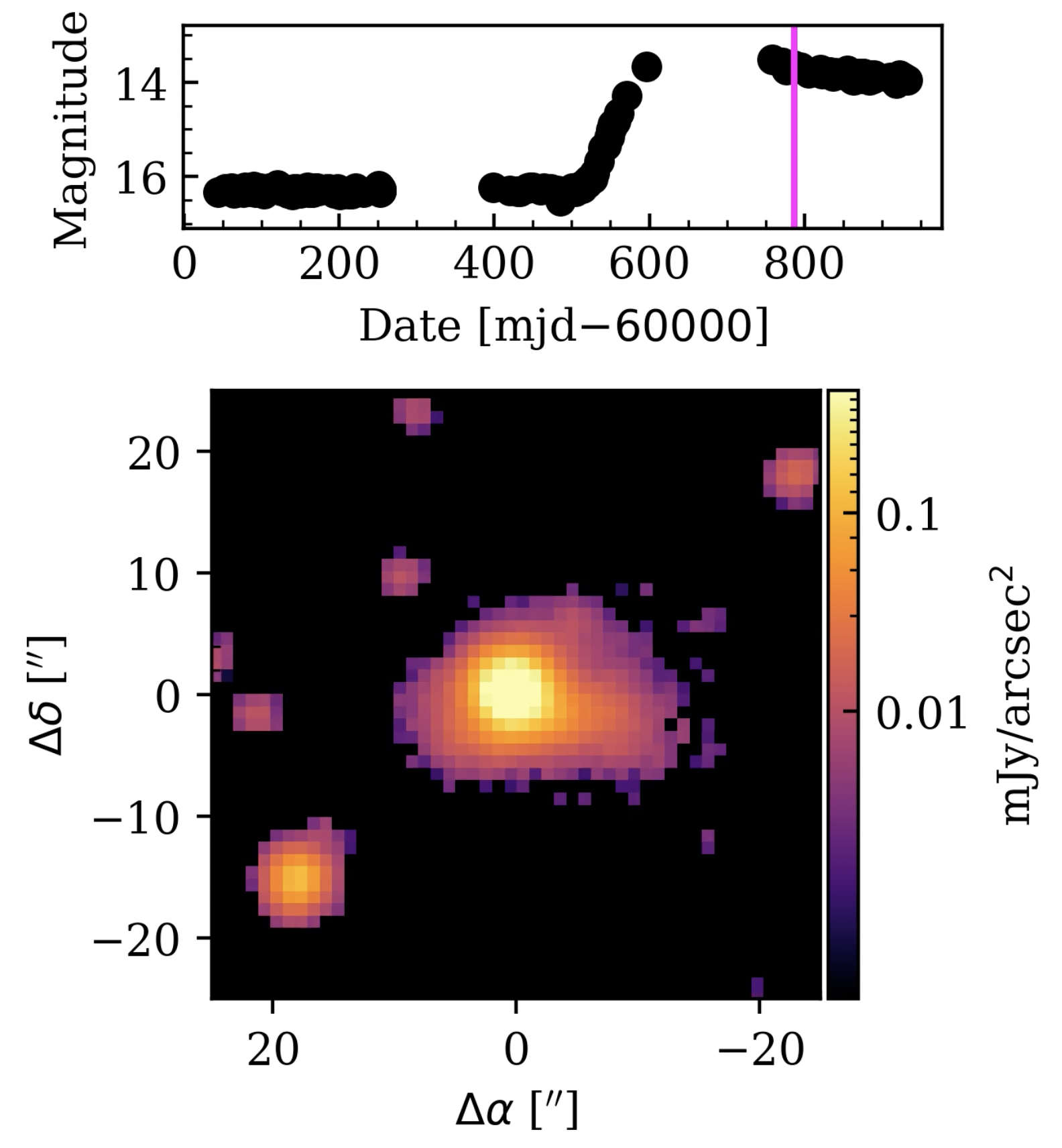}
    \caption{
    Still frame from an animation showing the
    ZTF r-band lightcurve and image of the post-outburst source. 
    There is a morphological change in the nebular environment as the source brightens.
    \textit{On-line version of the journal shows a 
    {\href{http://www.astro.caltech.edu/~lah/WTP19_ztfmovie_2panel.mp4}{movie}}. 
    }
    }
    \label{movie}
\end{figure}

\begin{appendix}
Figure~\ref{movie} shows $r$-band images from ZTF over the course of the outburst.
Only frames with seeing better than 3\arcsec.5 were used.
The images were calibrated to surface brightness using the reported zero points 
The images were calibrated to surface brightness units using the reported zero points and pixel scale from the image headers, then reprojected to a common WCS with standard North-up, East-left convention.

\end{appendix}

\bibliography{ms}{}
\bibliographystyle{aasjournal}

\end{document}